\documentstyle[aps,prl,epsf,multicol]{revtex}
\begin{document}
\draft
\title{An Interface View of Directed Sandpile Dynamics}
\author{Chun-Chung Chen and Marcel {den Nijs}}
\address{Department of Physics, University of Washington, P.O. Box 351560,
Seattle, Washington 98195-1560}
\date{March 30, 2001}
\maketitle
\begin{abstract}
We present a directed unloading sand box type avalanche model,
driven by slowly lowering the retaining wall at the  bottom of the slope.
The avalanche propagation in the two dimensional surface is related
to the space-time configurations of one dimensional
Kardar-Parisi-Zhang (KPZ) type interface growth dynamics.
We express the scaling exponents for the avalanche cluster
distributions into that framework. The numerical results
agree closely with KPZ scaling, but not perfectly.
\end{abstract}
\pacs{PACS numbers: 05.40.-a, 05.65.+b, 05.70.Np, 81.05.Rm}

\begin{multicols}{2}
\narrowtext

Avalanche phenomena are common in nature \cite{Bak}.
They are characterized by fast relaxation dynamics under a
slow driving force. Models that describe such dynamics, e.g., so-called
sandpile models, have been studied extensively for more than  a decade
following the work of Bak \emph{et al} \cite{Bak1987}.
Directed sandpile models are a special sub-class in which
relaxation follows a directional rule\cite{Dhar1989}, that is,
the propagation of active sites occurs only in one direction and
never backfires. The central issue in this type of research is
whether the dynamics is critical, such that the avalanche distribution
functions are scale invariant (power-law decay), and if so, whether
these scaling
properties are universal in the same sense as equilibrium critical phenomena.
Our understanding of these issues is still restricted.
There are only a few exactly soluble models, e.g.,
some deterministic Abelian directed sandpile models
\cite{Dhar1989}, but most insight is still limited to
numerical simulation data.
The evidence for scaling and universality
in other types of non-equilibrium dynamics is less ambiguous:
In surface growth (another example of intrinsic critical behavior)
several universality classes are well established;
e.g., Edwards-Wilkinson (EW) \cite{Edwards1982} and Kardar-Parisi-Zhang (KPZ)
\cite{Kardar1986} type surface growth;
population and catalysis type dynamics undergo absorbing state type
phase transitions with distinct universality classes like
directed percolation and directed Ising \cite{Kinzel1983,Marro,Noh1999}.

Efforts are under way to link avalanche dynamics to these better
understood dynamic phenomena. This ranges from mappings to driven interfaces
\cite{Sneppen1992,Paczuski1996,Lauritsen1999} and  directed percolation
\cite{Tadic1997,Vazquez1999} to using the concepts of
renormalization \cite{Pietronero1994,Vespignani1995} and universality classes
\cite{Ben-Hur1996,Mehta1996,Paczuski1996}.
In this letter we introduce a two dimensional (2D) directed avalanche model
in which the scaling exponents of the avalanche dynamics
are directly related to the exponents of KPZ
growth of a 1D interface.

\begin{figure}
\centerline{\epsfxsize=8cm \epsfbox{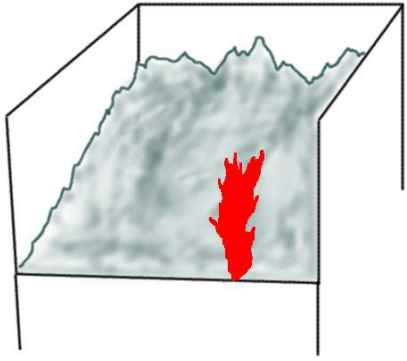}}
\vskip 10 true pt
\caption{Sand box with a slowly lowing retaining wall
\label{sand box picture}}
\end{figure}

The physical system we have in mind is a sand box with
a movable retaining wall to let out sand from the bottom
of the slope, see Fig.(\ref{sand box picture}).
The retaining wall is lowered very
slowly, such that grains tumble out sporadically forming distinct
avalanches instead of a continuous flow.
We model the sand surface by continuous height variables $h(x,y)$,
with respect to a 2D square lattice, which
is rotated over  $ 45^{\circ} $, meaning that
in the even (odd) $y$ rows $x$ takes only even (odd) integer values.
The retaining wall is placed at the $y=0$ boundary.
The slope is stabilized by the following constraint.
The surface particle in column $(x,y)$ is supported by
the two columns $(x\pm 1,y-1)$ just below it, and
must be lower than the lowest of the two increased by an amount $s_c$,
\begin{equation}
    \label{stability}
    h(x,y)\leq \min \left[ h(x+1,y-1),h(x-1,y-1)\right] +s_{c}.
   \end{equation}
An avalanche is triggered by selecting the
highest site, $(x_{i},0) $, at the $y=0$ wall boundary
(it is the $i$-th avalanche) and reducing its height
by a random amount, $0<\eta _{i}< s_{c}$.
This represents the  lowering of the retaining wall.
Next, all sites that violate the stability condition topple
according to the rule,
\begin{eqnarray} h(x,y) & \rightarrow & \min \left[
h(x+1,y-1),h(x-1,y-1)\right] \nonumber \\ & + & \eta
_{i}(x,y),\label{toppling}
\end{eqnarray}
where $0<\eta _{i}(x,y)\le s_c$ are uncorrelated random numbers.
This toppling continues until the whole system is stable again.
Since the toppling of a site in row $y$ can only effect the stability of
two sites immediately above it in row $y+1$,
the sites can be updated row by row starting from the $y=0$ boundary.

This process is idealized compared to a real unloading sand box
in the sense that the toppled grains drop out
without disturbing the already stabilized lower regions of the surface.
It is  possible to justify this as the low gravity or strong bond limit where
the falling sand does not gain
enough momentum to disturb the stabilized surface on its way out.
Without this idealization we loose the directed nature of the dynamics
and the following surface growth interpretation.

\begin{figure}
\centerline{\epsfxsize=8cm \epsfbox{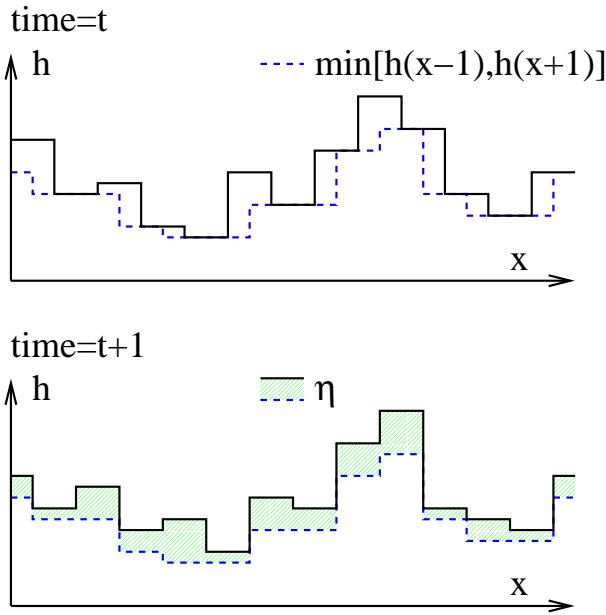}}
\vskip 10 true pt
\caption{KPZ type growth dynamics}
\label{growth dynamics}
\end{figure}

The row-by-row toppling sequence (\ref{toppling}) can be reinterpreted as
a dynamic rule for a 1D growing interface, in which the $y$
coordinate plays the role of time.
Each stable surface configuration represents a space-time
configuration of the 1D interface.
Imagine creating an initial stable surface configuration, before
the retaining wall starts to drop:
choose an arbitrary configuration with all $0<h(x,0)\le s_c$ in row $y=0$.
Next, apply Eq.(\ref{toppling}) to all sites in the
next row,  $y=1$, to create the next slice of the surface.
Repeat this  for all higher rows.
The configuration in every row is like a 1D interface evolving in
time $t=y$.
Fig.\ref{growth dynamics} illustrates how this interface propagates
during each time
step, $y\to y+1$. The upper panel shows the first half of the update
(the drawn to the dashed line). This is the deterministic part of
the propagation (the $\min[ \, ]$ operator) in Eq.(\ref{toppling}).
The lower panel illustrates the second half of the update,
where the heights increase by a random amount $0<\eta\le s_c$.
The first step removes material, and the second step deposits particles.

\begin{figure}
\centerline{\epsfxsize=8cm \epsfbox{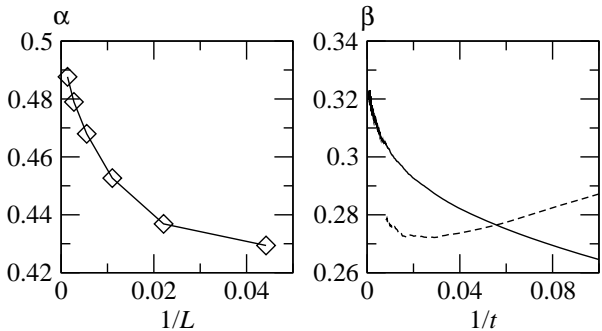}}
\vskip 10 true pt
\caption{Monte Carlo (MC) results for the global interface width:
(a) finite size ($L_x$)  estimates for the saturated surface width
exponent $\alpha$;
(b) finite time estimates for the transient surface width width
exponent $\beta$
from a flat initial configuration. The drawn line represent
conventional MC simulations
with an uncorrelated  world sheet ensemble, and the dashed line are data from
avalanche correlated MC runs.}
\label{showing of KPZ exponents}
\end{figure}

This type of interface dynamics almost certainly  belongs to the KPZ
universality class.
Eq.(\ref{toppling}) can be rewritten as
   \begin{eqnarray}
    h(x,t) & = & \frac{1}{2}\left[ h(x+1,t-1)+h(x-1,t-1)\right] \nonumber
    \\ & - & \frac{1}{2}\left| h(x+1,t-1)-h(x-1,t-1)\right| \nonumber \\ &
    + & \eta (x,t),\label{difference equation}
   \end{eqnarray}
which is a discrete version of the KPZ Langevin equation~\cite{Kardar1986},
   \begin{equation}
    \label{KPZ}
    \frac{\partial h}{\partial t}=\nabla ^{2}h-\frac{\lambda }{2}\left(
    \nabla h\right) ^{2}+\eta ,
   \end{equation}
To be absolutely sure, and also to make sure that
$\lambda$ is large enough such that corrections to scaling from
the EW point (at $\lambda=0$) do not interfere,
we checked numerically the behavior of the surface width  $\Delta (L,t)$,
defined as
$\Delta^2 \equiv \left\langle \left( h-\left\langle h\right\rangle
\right)^{2}\right\rangle$.
The scaling properties of KPZ growth are known exactly in 1+1 dimensions.
Starting from a flat surface at $y=0$, the width increases as
\( \Delta \sim t^{\beta } \) for \( 0\ll t\ll L^{z} \)
and saturates at
\( \Delta \sim L^{\alpha } \) for \( t\gg L^{z} \), with $L$ the
lattice size in the
$x$ direction.
The exponents are exactly equal to
$\alpha = \frac{1}{2}$, $\beta =\frac{1}{3}$ and $z=\alpha /\beta
=\frac{3}{2} $.
Our numerical results are shown in Fig.(\ref{showing of KPZ exponents}) and
are consistent  with the KPZ values.

Throughout this paper we present numerical results in terms of
finite size scaling  plots of effective exponents,
like $\alpha(L)$ in Fig.(\ref{showing of KPZ exponents})(a).
Global straight line fits to  log-log plots
of, e.g., $\Delta\sim L^\alpha$ are inaccurate.
$\alpha (L)$ is obtained from fitting the form $\Delta\sim L^\alpha$
to two nearby values of $L$. The approach to $L\to\infty$
in Fig.(\ref{showing of KPZ exponents}) is consistent with a leading
corrections to scaling exponent $y=-\frac{1}{2}$.

\begin{figure}
\centerline{\epsfxsize=8cm \epsfbox{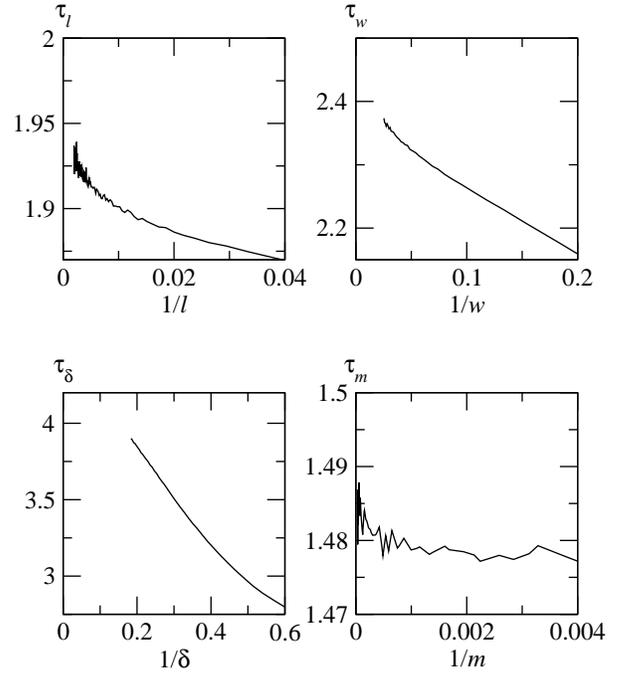}}
\vskip 10 true pt
\caption{Finite size scaling estimates of the avalanche distribution exponents,
$\tau _{l}$,
$\tau _{w}$,
$\tau _{\delta }$, and
$\tau _{m}$}
\label{tau exponents}
\end{figure}

The characteristic feature of self-organized criticality (SOC) is
the lack of typical avalanche length, width, depth, or mass scales.
The probability distributions follow power-laws, like $P_w\sim w^{-\tau_w}$
characterized by the scaling exponents, \( \tau _{l} \), \(
\tau _{w} \), \( \tau _{\delta } \) and \( \tau _{m} \).
Our numerical simulation results confirm the existence of scale invariance.
The  critical exponents converge well, see Fig.(\ref{tau exponents}).
The length $l$ is the maximum $y$ coordinate the avalanche reaches.
The width $w$ is the maximum departure in the \( x \) direction,
\(\left| x-x_{i}\right| \), from the trigger point \(x_{i} \).
The depth $\delta$ is the maximum height change, $h_{i}-h_{i-1}$,
caused by the avalanche, and the mass is the total
amount of sand carried off by the avalanche.

\begin{figure}
\centerline{\epsfxsize=8cm \epsfbox{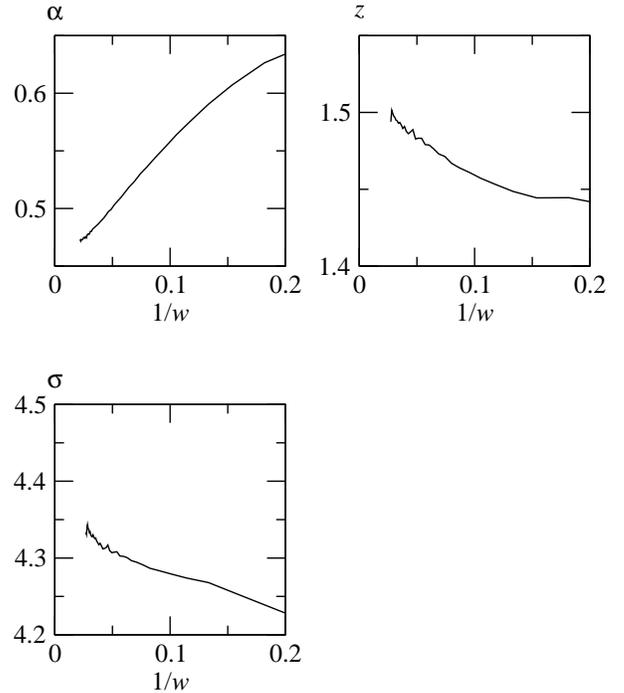}}
\vskip 10 true pt
\caption{Finite size scaling estimates for  $\alpha$, $z$ and
$\sigma$ derived from data in Fig.\ref{tau exponents}
using equation (\ref{alpha z sigma}).}
\label{azs exponents}
\end{figure}

The meta-distribution function, $P(l,w,\delta )$,
should obey the scaling form
   \begin{equation}\label{metaP}
    P(l,w,\delta )=b^{-\sigma}P(b^{-z}l,b^{-1}w,b^{-\alpha }\delta ).
   \end{equation}
with $b$ an arbitrary scale parameter.
The exponents, \( \sigma \), \( z \) and \( \alpha \),
are expected to be robust with respect to details of the dynamic rule,
and thus define the universality class.
Single parameter distributions, like $P_w\sim w^{-\tau_w}$,
follow by integrating out the other two parameters.
This implies the following expressions for the $\tau$ exponents
   \begin{equation}
    \tau _{l}=\frac{\sigma -1-\alpha }{z},\, \tau _{w}=\sigma -z-\alpha ,\,
    \tau _{\delta }=\frac{\sigma -1-z}{\alpha }.
   \end{equation}
or inverted,
   \begin{equation}
    \label{alpha z sigma}
    z=\frac{\tau _{w}-1}{\tau _{l}-1},\, \alpha =\frac{\tau _{w}-1}{\tau
    _{\delta }-1},\, \sigma =\tau _{w}+z+\alpha .
   \end{equation}
In Fig.(\ref{azs exponents}) we replot the numerical finite size
scaling estimates of the $\tau$ exponents in terms of $\alpha$, $z$,
and $\sigma$.

The values $z=1.52\pm 0.02$ and  $\alpha=0.46\pm 0.01$
are very close to those  of 1D KPZ growth.
Every stable slope configuration represents a possible world sheet of
a 1D KPZ type interface, and every avalanche the difference between
two such world sheets. Therefore it is natural to expect that
the length (depth) of the avalanche scales with the KPZ value
for $z$ ($\alpha$).

The distribution  of avalanche cluster sizes is the most
commonly studied  and experimentally the most
accessible property  of SOC.
Its exponent is linked to $\alpha$ and $z$ in the following manner.
At the start of the avalanche, the height of a boundary site (\( y=0 \))
is lowered on average by $\frac{1}{2} s_c$.
Thus, for a sandbox of width, $L_x$, the boundary row is lowered
by $\frac{1}{2}s_c$  after $L_x$ avalanches.
In the stationary state, the entire surface
moves down on average by $\frac{1}{2}s_c $ and the
average amount of removed sand is
equal to  $L_y\times \frac{1}{2} L_x s_c $.
Thus, the average mass of an avalanche must be equal to
   \begin{equation}
    \label{conservation}
    \left\langle m\right\rangle =\int dm\, mP_m(m)=\frac{1}{2}s_c L_y.
   \end{equation}
This is  analogous to conservation of current
in conventional deposition type avalanche systems
(see, e.g., \cite{Dhar1989}).
Assume that the avalanche clusters are compact,
i.e., that the sizes of holes of unaffected regions inside an
avalanche  do not scale with the avalanche size.
In that case, the mass scales as $ m\sim l\times w\times \delta$,
and we can use the meta-distribution function
to evaluate Eq.(\ref{conservation}), as
\begin{eqnarray}
\left\langle m\right\rangle
& \sim & \int ^{L_y}_0 dl\int_{0}^\infty dw\int_0^\infty d\delta
    ~~ lw\delta~P(l,w,\delta)\nonumber \\
& + & m_{L_y}\int _{L_y}^{\infty }dl\int_{0}^\infty dw\int_0^\infty d\delta
    ~~ P(l,w,\delta)
\label{average mass}
   \end{eqnarray}
This applies when the box is wide and deep enough that $L_y$ is the
only limiting factor to the avalanches. The first term accounts for
all avalanches that
fit inside the box, and the second term for the ones that reach the $L_y$ edge
and thus are prematurely terminated.
The first integral scales as \( L_y^{(-\sigma +2+2z+2\alpha
)/z} \) for large \( L_y \), while the second one scales as \( L_y^{(-\sigma
+1+z+\alpha )/z} \). We have assumed \( m\sim l\times w\times \delta
\) so \( m_{L_y}\sim L_y^{(1+z+\alpha )/z} \). The two terms in
Eq.(\ref{average mass}) scale in the same way, as
   \begin{equation}
    \label{scaling mass}
    \left\langle m\right\rangle \sim L_y^{(-\sigma +2+2z+2\alpha )/z}.
   \end{equation}
   Solving Eqs.(\ref{conservation}) and (\ref{scaling mass}) for \(
\sigma \) gives
   \begin{equation}
    \sigma =2+z+2\alpha
   \end{equation}
Numerically we find $\sigma =4.43\pm 0.05$, see Fig.(\ref{azs exponents}),
in agreement with the numerical values for $\alpha$ and $z$, and also with
their KPZ values (which yields $\sigma =9/2 $).

In conclusion, we introduced a directed unloading sand box model,
in which the stable slope configurations obey KPZ type scaling, and
the avalanches
represent the difference between two such KPZ world sheets.
The avalanche distribution exponents can be reformulated
in KPZ language, Eq.(\ref{alpha z sigma}),
and the numerical results agree with the KPZ values.

An intriguing issue for further study is whether the slight systematic
deviations from the KPZ values in Fig.(\ref{azs exponents}) are for real or
just an artifact of, e.g., finite size scaling
(the avalanches cover only a small part of the surface).
In KPZ dynamics the scaling exponents follow from an ensemble average over
independent MC runs, i.e., over a large set of independent
world sheets.
The avalanche dynamics performs this ensemble average in a correlated manner.
All subsequent world sheets are identical except for the avalanche area.
Correlated MC runs like this
might well change the scaling exponents (but only slightly apparently),
and this might prove  a key feature of avalanche type SOC.
The dashed curve in Fig.\ref{showing of KPZ exponents}(b) shows
initial results for the global surface roughness in such
avalanche correlated MC runs. The surface is noticeably rougher in amplitude
but the finite size scaling values of the exponent $\beta$ are
somewhat smaller (consistent with the values for $\alpha$ and $z$ in
Fig.\ref{azs exponents}).
Whether they will converge in the end to a  value
$\beta<\frac{1}{3}$ remains to be seen.

In this context it is noteworthy that
Eq.(\ref{alpha z sigma}) applies also to recent solutions to the
stochastic directed sandpile model (SDSM) of Paczuski {\it et.al.}
\cite{Paczuski2000} and Kloster {\it et.al.} \cite{Kloster2000} where
the avalanche dynamics is related
to Edwards-Wilkinson (EW) \cite{Edwards1982} surface growth, with
\( z=2 \) and \( \alpha =1/2 \); and also that
the exactly soluble directed sandpile model (DSM) of Dhar and Ramaswamy
satisfies Eq.(\ref{alpha z sigma}) with \( z=2 \) and \( \alpha =0 \).
However, that does not resolve the issue, because both models
are intrinsically simpler than ours, and the numerical accuracy
of SDSM in, e.g., Ref.\cite{Pastor-Satorras2000,Kloster2000} is not
sufficient to detect a small effect like this.

This research is supported by the National Science Foundation under grant
DMR-9985806.

\hfill{}

\end{multicols}
\end{document}